\documentclass[prb,preprint]{revtex4}
\usepackage{psfig}

\begin{document}

\title{Design of amino acid sequences to fold into C$_\alpha$--model proteins}
\author{A. Amatori$^{1,2}$, G. Tiana$^{1,2}$, L. Sutto$^{1,2}$, J.Ferkinghoff-Borg$^3$, A. Trovato$^4$ and R. A. Broglia$^{1,2,3}$}
\address{$^1$Department of Physics, University of Milano and} 
\address{$^2$INFN, sez. di Milano, via Celoria 16, 20133 Milano, Italy}
\address{$^3$The Niels Bohr Institute, Blegdamsvej 17, 2100 Copenhagen, Denmark}
\address{$^4$INFM and Dipartimento di Fisica "G. Galilei", Università di Padova, Via Marzolo 8, 35131 Padova, Italy}

\date{\today}

\begin{abstract}
In order to extend the results obtained with minimal lattice models to more realistic systems, we study a model where proteins are described as a chain of 20 kinds of structureless amino acids moving in a continuum space and interacting through a contact potential controlled by a 20$\times$20 quenched random matrix. The goal of the present work is to design and characterize amino acid sequences folding to the SH3 conformation, a 60--residues recognition domain common to many regulatory proteins. We show that a number of sequences can fold, starting from a random conformation, to within a distance root mean square deviation (dRMSD) of $2.6\AA$ from the native state. Good folders are those sequences displaying in the native conformation an energy lower than a sequence--independent threshold energy.
\end{abstract}

\maketitle

\section{Introduction}

A number of models have been studied in the last twenty years to describe the folding mechanism of single--domain proteins to their unique, biologically active native conformation. All--atom models with semi-empirical potentials \cite{gromacs,amber} provide a realistic description of proteins but are computationally too demanding to be useful to study their folding (cf. e.g. ref. \cite{duan}). A class of simplified models focuses on an accurate description of the geometry of the protein but employs minimal potential functions. It is the case of G\=o models \cite{go}, where the potential function sums a constant negative term for each native contact in the protein conformation. In these models the native conformation is by definition the ground state of the system and it is usually possible to perform extensive folding simulations. These models are used to study some features of selected proteins, such as the conformation properties of the transition state \cite{ts}. On the other hand the contribution of the different parts of the protein to its kinetics and thermodynamics is controlled mainly by the entropic term (the energetic term being trivial), and the frustration \cite{frustration} of proteins is underestimated. 

Lattice models use the opposite approach, accounting in a minimal way for the geometry of the protein chain, but focusing on the heterogeneity of the interactions \cite{lattice1,lattice2}. The protein is displayed as a chain of beads sitting on the vertices of a lattice interacting through a non--trivial contact matrix. These models account for the frustration of the system, allow the study of the amino acid sequences folding to a given model structure (and, consequently, of the effect of mutations, of the natural evolution of protein sequences, etc.) and are computationally quite economical. These models are used to understand the physical basis of the folding process, trying to answer questions such as what makes that a protein displays a low--entropy in its equilibrium state ?\cite{sh1},  what makes that a protein folds fast? \cite{jchemphys2,jchemphys3}, what differentiates a good folder from a random sequence ? \cite{lattice2}, etc. On the other hand, lattice models cannot describe the fold of real proteins with its richness of secondary structures and motifs.

The importance of the heterogeneity in the interaction between amino acids relies on the fact that analytical calculations made on random sequences with a replica approach have shown \cite{sh1} that there is a threshold degree of heterogeneity which separates two qualitatively different behaviours of the system. For a low degree of heterogeneity any model chain behaves essentially as a globular homopolymer, displaying at any temperature an equilibrium state populated of many (i.e., exponentially many with respect to the chain length) different conformations. Within this context it is not possible to find protein--like sequences with a unique native state. At high heterogeneity, sequences with few dominant conformations appear and a fraction of them have a unique equilibrium conformation \cite{sh2}. These are the candidates to the role of good folders.

Simulation studies based on lattice models have highlighted (see, among others, ref. \cite{klimov}) a simple energetic criterion to distinguish good from bad folders; a sequence will fold to a given native conformation if it displays, on that conformation, an energy $E_N$ lower than a threshold energy $E_c$, energy which only depends on the statistical moments of the interaction matrix and on the length of the chain \cite{sh1}. The physical meaning of $E_c$ is that of being the lowest conformational energy a random sequence can have, energy which has a well--defined value as a consequence of the frustration of the system \cite{derrida}. This condition ($E_N<E_c$) goes further than to ensure the thermodynamical unicity of the native state. In fact, it is also at the basis of the kinetic ability the protein has to reach the native state on short call \cite{jchemphys2,jchemphys3}. 

An energy minimization of the sequence, keeping the protein conformation fixed, to energies below $E_c$ is thus a practical algorithm to design good folders. This procedure has been applied to the design of lattice model proteins \cite{sh1}. A more efficient and thermodynamically rigorous method consists in optimizing the conformational free energy of the sequence in the protein conformation, thus taking into account also the competing conformations. This method has been introduced in refs. \cite{deutsch1,deutsch2} and further developed in refs. \cite{mich1,irbaek,mich2,mich3}.

The purpose of the present work is to build a model which is more realistic of both G\=o and lattice models, including continuous degrees of freedom as well as a non--trivial potential function. We show that this model allows sequences to have a unique, stable and kinetically accessible native state, and that such sequences obey the same energetic requirement as lattice--model sequences do. With the help of this model we will investigate the folding of selected sequences into the SH3 domain.

\section{The protein model} \label{sect_2}

The model we investigate describes a protein as an inextensible chain, where amino acids are represented by spherical beads centered around the C$^\alpha$-atom, thus allowing a realistic accounting for the protein backbone (cf. Fig. \ref{fig_native}). Each of the 20 types $\sigma$ of amino acids is characterized by a hard core radius $R^{HC}(\sigma)$. The values of $R^{HC}(\sigma)$, which range from $2.25 \AA$ to  $3.39 \AA$,  are listed in Table \ref{tab_aa}. The bond angles are limited within the interval between $\pi/2$ and $0.8\pi$ so as to give some amount of stiffness to the polypetide chain.

The potential energy of the protein depends on the positions $\{r_i\}$ and sequence $\{\sigma_i\}$ of amino acids according to 
\begin{equation}
U(\{r_i\},\{\sigma_i\})=\sum_{i<j+1}B(\sigma_i,\sigma_j)\theta\left(R(\sigma_i,\sigma_j)-|r_i-r_j|\right),
\end{equation}
where $\theta$ is Heaviside's step function, $R(\sigma_i,\sigma_j)\equiv k\cdot(R^{HC}(\sigma_i)+R^{HC}(\sigma_j))$ is the range of interaction which depends on the kind of amino acids proportionally to their hard--core radius ($k= 0.721$) and $B_{\sigma\pi}$ is the interaction energy between an amino acid of kind $\sigma$ and one of kind $\pi$. The matrix $B_{\sigma\pi}$ is built out of quenched \footnote{Quenched in the sense that they are generated at the beginning of the investigation and maintained fixed.} random numbers taken from a Gaussian distribution with mean $B_0=0.25$ and standard deviation $\sigma_B=0.52$ (in arbitrary units). A slighlty positive mean contact energy has been chosen because it leads to sequences with better folding properties than those associated with a contact matrix displaying $B_0\leq 0$. 

The rationale behind the model is that the key ingredient to make a protein fold is the heterogeneity existing among the different amino acids. We account for this heterogeneity by both varying the size of the amino acid as well as the contact energy acting among them. Because no simple energy function capable of folding real protein sequences to their native conformation is yet known (see e.g. ref. \cite{karplus}), we shall use in the model calculations an interaction potential parameterized by a random matrix. This choice has the advantage to make the model results quite general. On the other hand, it has the drawback that the labels of amino acids (A, C, etc.) are merely nominal. Since we are interested in the general aspects of the physical mechanism of protein folding and not in the detailed chemistry of particular sequences, this drawback is of no consequence for the present investigation.

An important ingredient of the model is a constrain on the total number of contacts each residue can build. This constrain reflects (together with the hard core radius) the size of the given amino acid. Simulations performed without such a constrain led to a collapse of the chain to unrealistically small sizes. The effect cannot be avoided be simply increasing the hard core radius, since this value is limited from above by the average distance between two amino acids along the chain ($\approx 3.8\AA$). This is a limit any model which pictures an amino acid as a sphere will display. In keeping with the discussion carried out above, we assign to each type of amino acid a quenched random number $n_{max}(\sigma)$ ranging from 0 to 5, which represents the maximum number of possible contacts the amino acid can make (cf. Table \ref{tab_aa}). 

In order to design sequences with a specified energy $E_{targ}$ on the SH3 target conformation (Fig. \ref{fig_native}(a)) we perform a sampling of the space of sequences. Starting from a random sequence displaying an even concentration of amino acids (operatively, we have mantained that of the wild--type Src SH3, although this choice has not any deep meaning, due to the merely nominal character of the amino acid letters), we keep the conformation fixed and perform swap moves among the amino acids, so that their relative frequency remains constant. Consequently, the average and standard deviation of the contact energy matrix weighted by this frequency remain equal to that associated with the (unweighted) $20\times 20$ matrix. Examples of sequences obtained with this algorithm are listed in Table \ref{tab_seq}. We note that, although wild--type sequences do display some amount of amino acid repetitions, the designed sequences display unrealistic repeats of amino acids of the same kind (see Table \ref{tab_seq}). This is an artifact of the simplified spherical geometry of model residues, which causes correlations among consecutive sites. In fact, whenever the $j$-$th$ residue ($j$ being the site index along the chain) interacts with residue $j^*$, then residues $j-1$ and $j+1$ are likely to lie within the interaction range of $j^*$ as well. This effect is smaller in real proteins because of the complicated geometry of sidechains. Since this artifact of the model does not put obstacles in the designing of good folders, we will postpone the solution of this problem to a future work, where the folding sequences will be analyzed from a bioinformatic point of view.

\section{Conformational analysis of the lowest--energy sequence}

The first sequence analyzed is that displaying the lowest energy $E_{targ}=-37.80$ on the SH3 conformation, and labeled $s_1$ in Table \ref{tab_seq}. We have performed Monte Carlo simulations (each of $10^9$ steps) in conformational space at fixed temperatures, ranging from the value $0.05$ to $1.0$, starting each time from a random conformation. In 10 simulations at $T=0.12$ the sequence $s_1$ finds each time an energy minimum displaying an dRMSD \footnote{We define dRMSD as the root of the mean square difference between the inter--residue distance in the given conformation and in the native state, calculated over all pairs of residues. As a reference, note that the dRMSD of a random conformation displays a dRMSD of the order of $25\AA$ to the native conformation of SH3. On the other hand, the meaning of a dRMSD of $2.6\AA$ can be appreciated from Fig. 1.} to the SH3 target conformation (Fig. \ref{fig_native}(a)) smaller than $3.3\AA$ and with the relative number $q$ of native attractive contacts larger than $0.80$. The minimum energy, that is the ground state energy found in the 10 runs is $E_{gs}=-46.96$ and is associated with a conformation displaying a dRMSD of $2.6\AA$ and a $q$ of 0.85 (Fig. \ref{fig_native}(b)). No conformation dissimilar from the native conformation and displaying a energy lower than $-46.96$ is found in Monte Carlo simulations at any temperature. 

The ground state energy is smaller than the energy $E_{targ}=-37.80$ found in the sequence minimization. This is because the ground state conformation displays $N_c=74$ contacts, while this number is 67 for the target conformation and each of the 7 new contacts has in average an energy of $-1.3$. Very--low temperature simulations ($T=0.01$) starting from the target conformation converge rapidly into the ground state conformation, thus indicating that the two conformations belong to the same basin of attraction. The reason why the target configuration is not exactly the ground state conformation of the system is that, during sequence optimization, the native conformation is kept strictly fixed. If one were interested in making the target and the ground state conformations coincide exactly, one should allow some degree of conformational relaxation during sequence minimization. That is, perform a minimization in the crossproduct space of sequences and configurations.  

In Fig. \ref{fig_dyn} we display the dRMSD and the fraction\footnote{Only attractive contacts are counted and the same definition of contact as in the potential function (Eq. (1)) is used to calculate $q$.} of native (attractive) contacts $q$ associated with sequence s$_1$ as functions of the number of Monte Carlo steps. As in the case of G\=o-- \cite{go2} and of the lattice--model designed sequences \cite{lattice2}, the protein wanders between unfolded states (dRMSD $\approx 6\AA$) until it suddenly finds the native energy basin (dRMSD$<3\AA$).  Because the conformational move implemented in the Monte Carlo simulation is a small flip of a random--selected amino acid, one could also view such a simulation as an approximation to the dynamical trajectory \cite{rey}. Note also that, unlike G\=o models \cite{go}, in the potential function used in these simulations there is no information concerning the native conformation of the protein. 

During the simulations one observes a number of transitions from the folded to the unfolded conformation and vice versa (Fig. \ref{fig_dyn}). These transitions are reflected by changes in energy, the mean energy difference between the two states being about 2 energy units (i.e., $\approx 20\;kT$). On the other hand, it is difficult to distinguish between folded and unfolded states from the fraction of native attractive contacts $q$ alone (lower panel in Fig. \ref{fig_dyn}). This fact can be regarded as an indication that not all contacts partecipate on equal footing to the stability of the native conformation.

The thermodynamics of sequence $s_1$ has been studied by means of a multicanonical sampling algorithm \cite{jesper}.  The probability distribution at $T=0.12$ as a function of energy and dRMSD is shown in Fig. \ref{fig_PdiE}. The plot shows a clear two--state behaviour. The centroids of the two peaks are characterized by the values $E=-42$, dRMSD=$3.1$ (native state) and $E=-41$, dRMSD=$6.3$ (unfolded state), respectively. The fact that at this temperature the volumes defined by the two peaks are equal qualifies $T=0.12$ as the folding temperature. Note that the energetic difference between the two peaks is only $\approx 8\;kT$, while the energy fluctuations amount to $\approx 25\;kT$, and thus the energy distributions overlap consistently. Consequently, it is difficult to identify the two states only from the energy distribution of sequence in conformational space. Anyway, the plot shows that the lowest energy that the unfolded state can reach is $-42.5$. 

The specific heat associated with sequence s$_1$ is displayed in Fig. \ref{cv} and shows two major peaks at $T=0.12$ and $T=0.34$. These peaks indicate cooperative kinds of transitions and will be further investigated in a successive work. 

In Fig. \ref{fig_dRMSD} the mean dRMSD of the s$_1$ and of the s$_9$ (random sequence) are displayed as a function of the conformational energy. For energies larger than $-41$ the mean dRMSD of the designed sequence $s_1$ is very similar to that of non--designed sequence $s_9$, displaying a wide plateau at dRMSD 5.7$\AA$, corresponding to unfolded conformations. The designed sequence $s_1$, on the other hand, displays a transition around $-41.5$ to dRMSD $2.8\AA$. This allows us to define the native basin as the set of conformations displaying a dRMSD lower than $4.2\AA$, that is the midpoint of the transition, also consistently with the fluctuations observed in Fig. \ref{fig_dyn} and with the transition state of Fig. \ref{fig_PdiE}.

\section{Folding properties of different sequences}

The conformational analysis has been repeated for other 9 sequences displaying various values of $E_{targ}$ (cf. Table \ref{tab_seq}). For every sequence the average dRMSD obtained in 10 independent Monte Carlo simulations as a function of $E_{targ}$ is reported in Fig. \ref{fig_seq}. One can observe a monotonic behaviour up to $E_{targ}\approx -35$, where the dRMSD assumes a value of $4.2\AA$, which we have defined in the previous Section as the threshold between native and unfolded state. Moreover, the dRMSD associated with the ground state conformation is also listed in Table \ref{tab_seq}. Sequences displaying $E_{targ}$ above $-35$ are neither able to find during the simulations conformations similar to the target one, nor populate a set of structurally similar conformations (cf. inset of Fig. \ref{fig_entropy}). This result allows us to obtain, within the framework of the model introduced in Sect \ref{sect_2} energetic criterion to design protein--like sequences. In fact, good folders onto the SH3 domain are those sequences displaying an energy $E_{targ}<E_{targ}^c$, where $E_{targ}^c=-35$.

While the dependence of the folding properties of a sequence on $E_{targ}$ are quite clear, the dependence of these properties on $E_{gs}$ are less well--defined (cf. Table \ref{tab_seq}). In fact, while sequences with $E_{gs}$ well below $-44$  are guaranteed to fold (like the case of $s_1$, $s_2$ and $s_3$) it is difficult to assess the behaviour of sequences displaying values of  $E_{gs}$ around $-44$ (see, e.g., s$_5$ and s$_6$). The problem arises mainly because the minimum energy conformations associated with these sequences display different numbers of contacts ($N_c$ varies between 67 and 74), and consequently the system can gain or loose an amount of energy of the order of some $kT$ with ease. Anyway, for the purpose of designing a good folder into a given three--dimensional conformation what matters is the $E_{targ}$ criterion.

In Fig. \ref{fig_entropy}, the conformational entropy of three protein--like sequences, namely $s_1$, $s_2$ and $s_3$ are shown as a function of energy. For reference, the entropy of the non--folding sequences $s_8$, $s_9$ and $s_{10}$ are also plotted. The entropy of high--energy states ($E>-20$) is very similar for all sequences except the purely random ones ($s_9$ and $s_{10}$). This is consistent with the idea that in high--energy conformations contact energies can be regarded as random, any specific effect of the individual sequence being lost. In fact, the high--energy part of the plot can be well approximated by means of the random energy model \cite{derrida}, where the total energy $E$ of the system is described as the sum of $N_c$ uncorrelated stochastic contact energies, $N_c$ being the typical number of contacts in a globular conformation. The resulting entropy is
\begin{equation}
S(E)=S_0-\frac{(E-N_c B_0)^2}{2N_c\sigma_B^2},
\label{s_e}
\end{equation}
where $S_0$ sets the zero of the entropy, $B_0=0.25$ and $\sigma_B=0.52$ are the mean and standard deviation of the interaction matrix. The curve described by Eq. (\ref{s_e}) is plotted with dotts in Fig. \ref{fig_entropy}, fitting the values of $S_0$ and of $N_c$ ($=29$) to the high temperatures part of the curves obtained from the simulations. Below energy $E\approx-20$ the entropy of these sequences is influenced by the specificity of the sequence, as evinced by the departing of the curves from Eq. (\ref{s_e}).

The random sequences s$_9$ and s$_{10}$, on the other hand, display as expected an entropy function which is qualitatively different from the folding sequences. Furthermore, this cannot be fitted satisfactorly with Eq. (\ref{s_e}). This is somewhat unexpected, since a random sequence should be described better than a designed sequence by the random energy model.

Moreover, the non--designed sequences not only display low--energy conformations structurally dissimilar from the target conformation, but these conformations are also dissimilar among themselves. The inset of Fig. \ref{fig_entropy} shows the distribution of dRMSD for a good and a bad folder, calculated pairwise in a sample of 20 conformations displaying $E_{gs}<E<E_{gs}+10\;kT$. The centroid of the distribution associated with sequence s$_9$ lies at $10\;\AA$ (dashed curve), indicating that the associated conformations are structurally very different.

As expected, the ground state energy of random sequences is higher than that of good folders (e.g. $E_{gs}=-42.36$ and $E_{gs}=-39.36$ for s$_9$ and s$_{10}$, respectively). Consistently with the results of lattice models \cite{lattice2}, the mean ground state energy of random sequences is approximately equal to the lowest energy of the unfolded state of a good folder ($\approx -42$, see Fig. \ref{fig_PdiE}), and we shall call $E_c$ this energy.

Consistently with these findings, a sufficient (but not necessary; cf. e.g. sequence $s_4$) condition for any sequence to be a good folder is that it displays a ground state energy $E_{gs}$ much lower than $E_c$ (cf. Table \ref{tab_seq}). The reason is that, since $E_c$ is essentially sequence--independent, if a sequence displays $E_{gs}\ll E_c$, then its conformational ground state has not to compete with the sea of unfolded conformations. Differently from the case of lattice models, this result is only partially predicitive. While for lattice--model sequences the folding requirement is just $E_{gs}<E_c$ \cite{jchemphys2}, in the present model it is important, although not well--defined, the "much lower" requirement. The reason for this difference is that in the present model there are consistent fluctuations in the number of contacts, which produce fluctuations in the energy. In a lattice model, due to the discreteness of the degrees of freedom, this effect is much smaller, and the overall behaviour is consequently clearer. On the other hand, one can easily distinguish good from bad folding sequences on the basis of the target energy $E_{targ}$ which, being calculated on a fixed conformation, does not undergo such fluctuations.

There are other features which, although not being usable for the design, set a physically clear difference between folding and non--folding sequences. First, the density of states of designed sequences at low and intermediate energy is much higher than for random ones (see Fig. \ref{fig_entropy}). That is, it is higher if the folding sequence is better designed. This is equivalent to state that the conformational accessibility of the ground state of well designed sequences is greater than for random or bad designed ones. In other words, asking for a deep funnel to be carved in the energy landscape ensures it to be also a wide one, consistently with the findings of ref. \cite{amos}. The second discriminant between `good' and `bad' folders is clearly seen in the fraction of native attractive contacts vs. energy curve (Fig. \ref{fig_QdiE}). The linearity of such curve for well designed sequences is, on the one hand, a striking confirmation that simple topology-based models (G\=o-model), which assume the energy gain to be proportional to the fraction of native attractive contacts, do indeed capture the basic feature of the energy landscape for a `good' folding sequence, i.e. the existence of a funnel towards the native state. On the other hand, it shows that our simple model is able to reproduce such crucial feature without any `a priori' knowledge of the native state. Random or `bad' folding sequences instead fail in creating the proper folding funnel.  Note that both features can be easily appreciated only in the microcanical ensemble by looking at the behaviour of entropy (fraction of native attractive contacts) as a function of energy.

The energy distribution per site of s$_1$ in the target conformation is also typical of good folders, as found in the case of lattice models \cite{jchemphys1}, the energy being concentrated mainly in few "hot" sites (cf. inset to Fig. \ref{fig_QdiE}). On the contrary, the stabilization energy of a random sequence is, as expected, evenly distributed over the ground state of the protein.

\section{Conclusions}

In the case of simple lattice models, the thermodynamics of heteropolymers is reasonably well--understood, and there is an efficient algorithm to design folding sequences given a target conformation and an interaction matrix. We have studied a model with continuous degrees of freedom, showing that it is possible to design 20--letters sequences which fold stably and fast to a given conformation, without that the potential function contains any information about the target conformation. A key ingredient of the model is a constrain on the total number of contacts that each amino acid can build, which reflects geometric features of the amino acid neglected by a spherical--bead model. By means of such a model, it is possible to conclude that good folder sequences are those displaying on the target conformation an energy lower than a sequence--independent threshold.

\clearpage
\newpage

\begin{table}
\begin{tabular}{|c|c|c|c|c|c|}
\hline
 & $R^{HC}$ & $n_{max}$ & & $R^{HC}$ & $n_{max}$ \\\hline
A       &  2.524     &   5    & M     & 3.099      &   5    \\
C       &  2.743     &   5    & N     & 2.845      &   2    \\
D       &  2.795     &   3    & P     & 2.790      &   4    \\
E       &  2.968     &   4    & Q     & 3.013      &   2    \\
F       &  3.188     &   2    & R     & 3.287      &   1    \\
G       &  2.258     &   1    & S     & 2.597      &   3    \\
H       &  3.048     &   4    & T     & 2.816      &   0    \\
I       &  3.099     &   4    & V     & 2.922      &   2    \\
K       &  3.188     &   2    & W     & 3.395      &   2    \\
L       &  3.099     &   5    & Y     & 3.234      &   4    \\
\hline
\end{tabular}
\caption{The features of the amino acids.}
\label{tab_aa}
\end{table}

\begin{table}
\begin{tabular}{|c|c|c|c|l|}
\hline
label & $E_{targ}$ & $E_{gs}$ & dRMSD$[\AA]$ & \\\hline
$s_1   $   & -37.80 &-46.96 & 2.6 &    {\tiny GLLLLAANNWWVTRTDEEKKDYVSSSSDDTQTGGYNIEGLIFFRQVVPPEAHTYYSSSTT}\\
$s_2   $   & -37.27 &-46.22 & 3.0 &    {\tiny GLLLLEEEGWWNGTTVYYKFDESPDSSSDTNGVTNYYVLFITRRVQQAAADHTPISSSKT}\\
$s_3   $   & -36.20 &-45.58 & 3.2 &    {\tiny NKSAAAHQPERFTTVSSSEEPIYEVLLNWTYTTTRDSDSDDKFGWGGLLLQGTIYVNSVY}\\
$s_4   $   & -35.53 &-44.45 & 3.4 &    {\tiny QQHAASSSDDSDVFTVPPLGNLTNYYGIITKTTWLLFEGGAYTRNVDEEESSTLSVKYRW}\\
$s_5   $   & -34.85 &-44.55 & 4.0 &    {\tiny GDSAAAHQPERWWTTSSSEEPIYEVLLNVTTTFTRDVDSSDKVGFNGLLLQGTIYYNSKY}\\
$s_6   $   & -34.30 &-44.31 & 4.7 &    {\tiny QWAAHEEEDYRNFGTSSSYQGPGINSSFKTGYTTVDSDSLATRVVVDLLLILWEPKNYTT}\\
$s_7   $   & -33.65 &-44.22 & 4.8 &    {\tiny SGLNLEEPGKKYFRRTAAWFVEGSDSSVGTTTTNQHQTALLLWVSDDYYYIIVEPDSSTN}\\
$s_8   $   & -23.67 &-42.09 & 4.8 &    {\tiny DSSSSEERDIFYTTTWYYQQGPLNSLLLGTVKTVDDIYSSAKTRWVGAAHGPTEEFNLVN}\\ \hline
$s_9   $   &  -4.51 &-42.36 & 5.5 &    {\tiny NLILYEKLDNRFNKWWFLADSSPASGQVDRTTSTVSSTQEHTTYEEYVSGLGTIPDAVGY}\\
$s_{10}$   &  +8.26 &-39.36 & 5.9 &    {\tiny EYLSVIKTEDPKQSEYPSWLSEFFLLTIATGNTLYYDGVHAVTSSRNSGGDAVRNDTTWQ}\\
\hline
\end{tabular}
\caption{Sequences with selected energies $E_{targ}$ on the SH3 target conformation displayed in Fig. \protect\ref{fig_native}. The reported dRMSD is that of the ground--state conformation. Sequences s$_9$ and s$_{10}$ are purely random.}
\label{tab_seq}
\end{table}

\newpage
\clearpage

\begin{figure}
\caption{(a) The native structure in a $C_\alpha$ representation of SRC SH3 as obtained by crystallographic experiments (pdb code 1FMK) and (b) the minimum energy structure of the sequence $s_1$, corresponding to a dRMSD of $2.6\AA$. }
\label{fig_native}
\end{figure}

\begin{figure}
\caption{The dRMSD (above), the energy (middle) and the fraction of native attractive contacts $q$ (below) as a function of the number of steps for a simulation of sequence $s_1$ starting from a random conformation at $T=0.120$.}
\label{fig_dyn}
\end{figure}

\begin{figure}
\caption{Probability distribution as function of energy and dRMSD for sequence $s_1$ at temperature 0.120.}
\label{fig_PdiE}
\end{figure}

\begin{figure}
\caption{The specific heat $C_v$ as a function of temperature for the sequence $s_1$.}
\label{cv}
\end{figure}

\begin{figure}
\caption{The average dRMSD as a function of energy, calculated for sequence $s_1$ (above) and $s_7$ (below). The error bars indicate the dRMSD standard deviation.}
\label{fig_dRMSD}
\end{figure}

\begin{figure}
\label{fig_seq}
\end{figure}

\begin{figure}
\caption{The conformational entropy as a function of energy for some of the sequences listed in Table \protect\ref{tab_seq}. Solid curves indicate folding sequences, dashed curves non--folding sequences. The dotted curves are the prediction of the random energy model. In the inset, the distribution of dRMSD for low--energy conformation sampled with sequence s$_1$ (solid curve) and s$_8$ (dashed curve).}
\label{fig_entropy}
\end{figure}

\begin{figure}
\caption{Fraction of native attractive contacts $q$ as function of energy for sequences $s_1$ (straight line), $s_8$ (dashed) and $s_9$ (dotted), representing respectively a good folder, a bad folder and a random sequence. In the inset, the distribution of stabilization energy among the residues in the target conformation of s$_1$.}
\label{fig_QdiE}
\end{figure}

\newpage
\clearpage

\begin{figure}
\centerline{\psfig{file=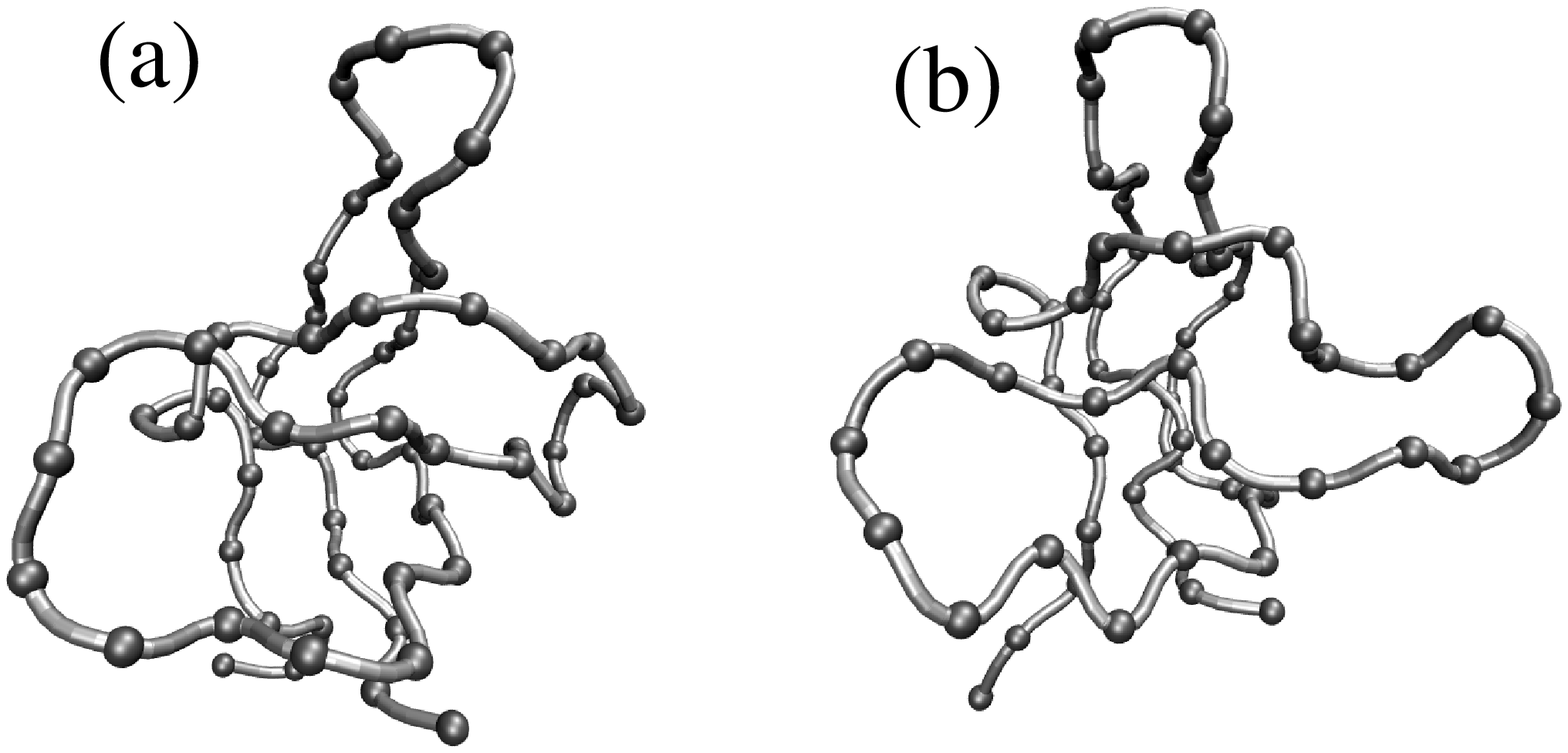,width=12cm}}
\end{figure}

\newpage
\clearpage

\begin{figure}
\centerline{\psfig{file=dyn.eps,width=12cm}}
\end{figure}

\newpage
\clearpage

\begin{figure}
\centerline{\psfig{file=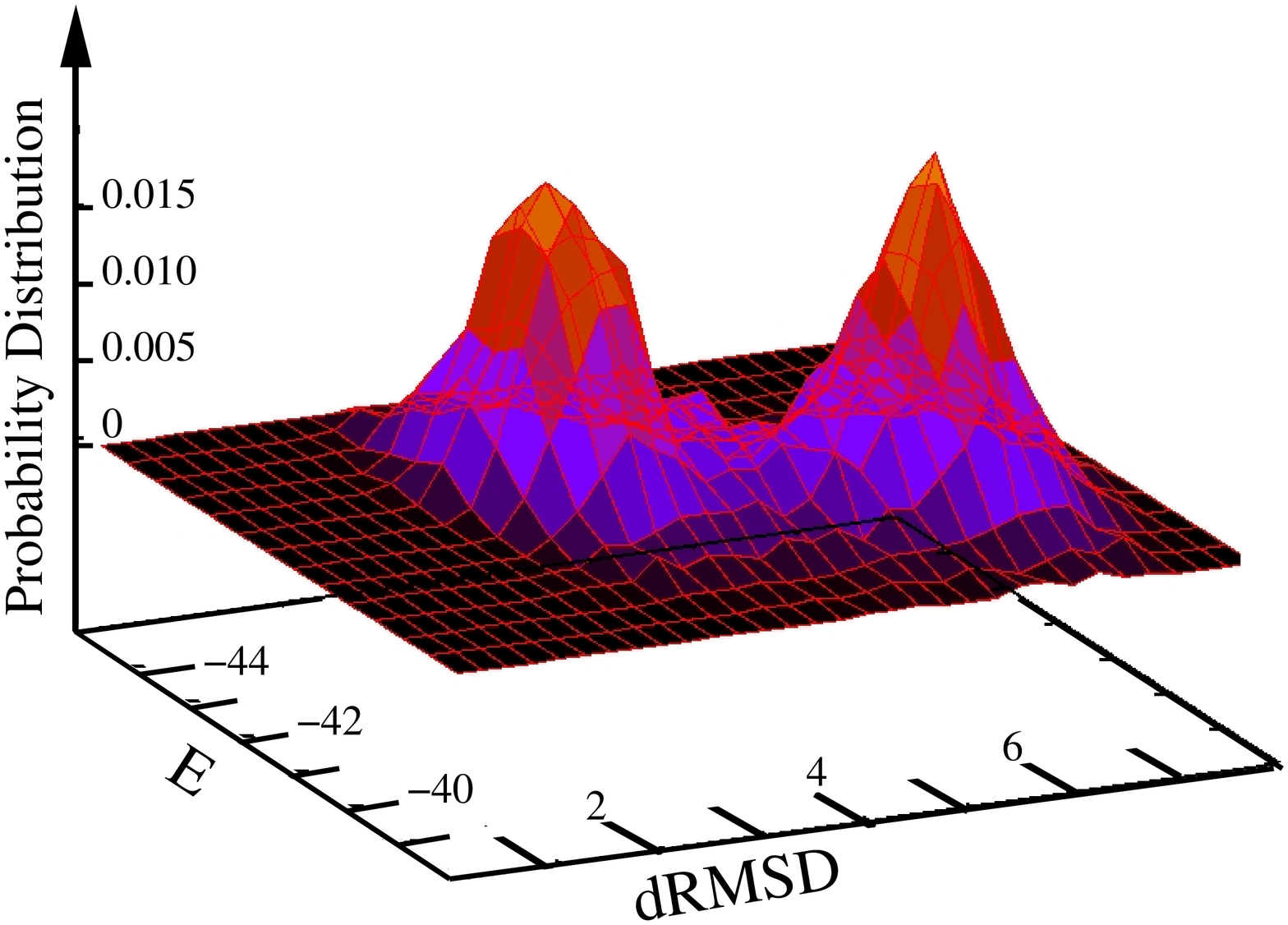,width=12cm}}
\end{figure}

\newpage
\clearpage

\begin{figure}
\centerline{\psfig{file=cv.eps,width=12cm}}
\end{figure}

\newpage
\clearpage

\begin{figure}
\centerline{\psfig{file=rmsd.eps,width=12cm}}
\end{figure}

\newpage
\clearpage

\begin{figure}
\centerline{\psfig{file=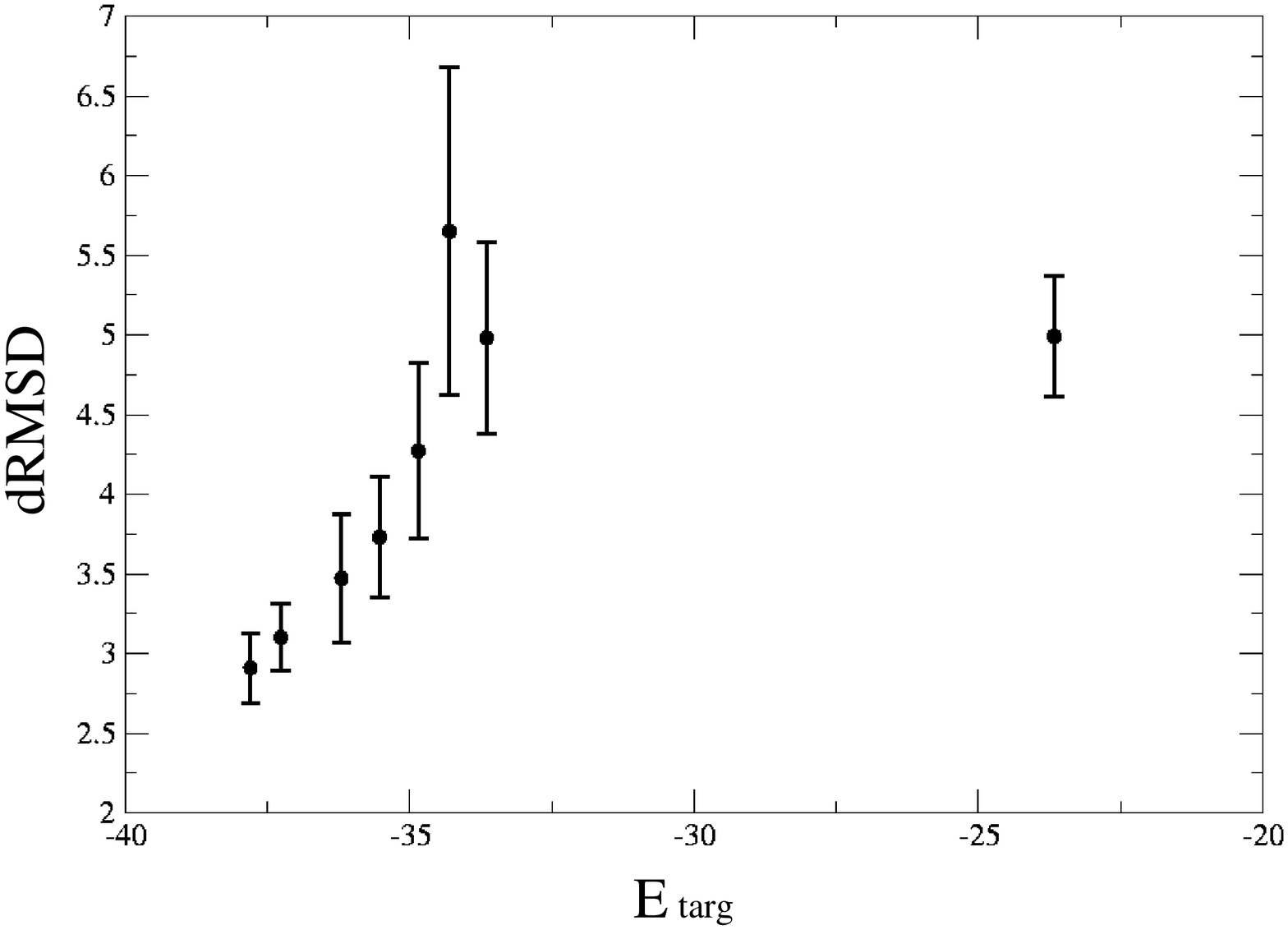,width=12cm}}
\end{figure}

\newpage
\clearpage

\begin{figure}
\centerline{\psfig{file=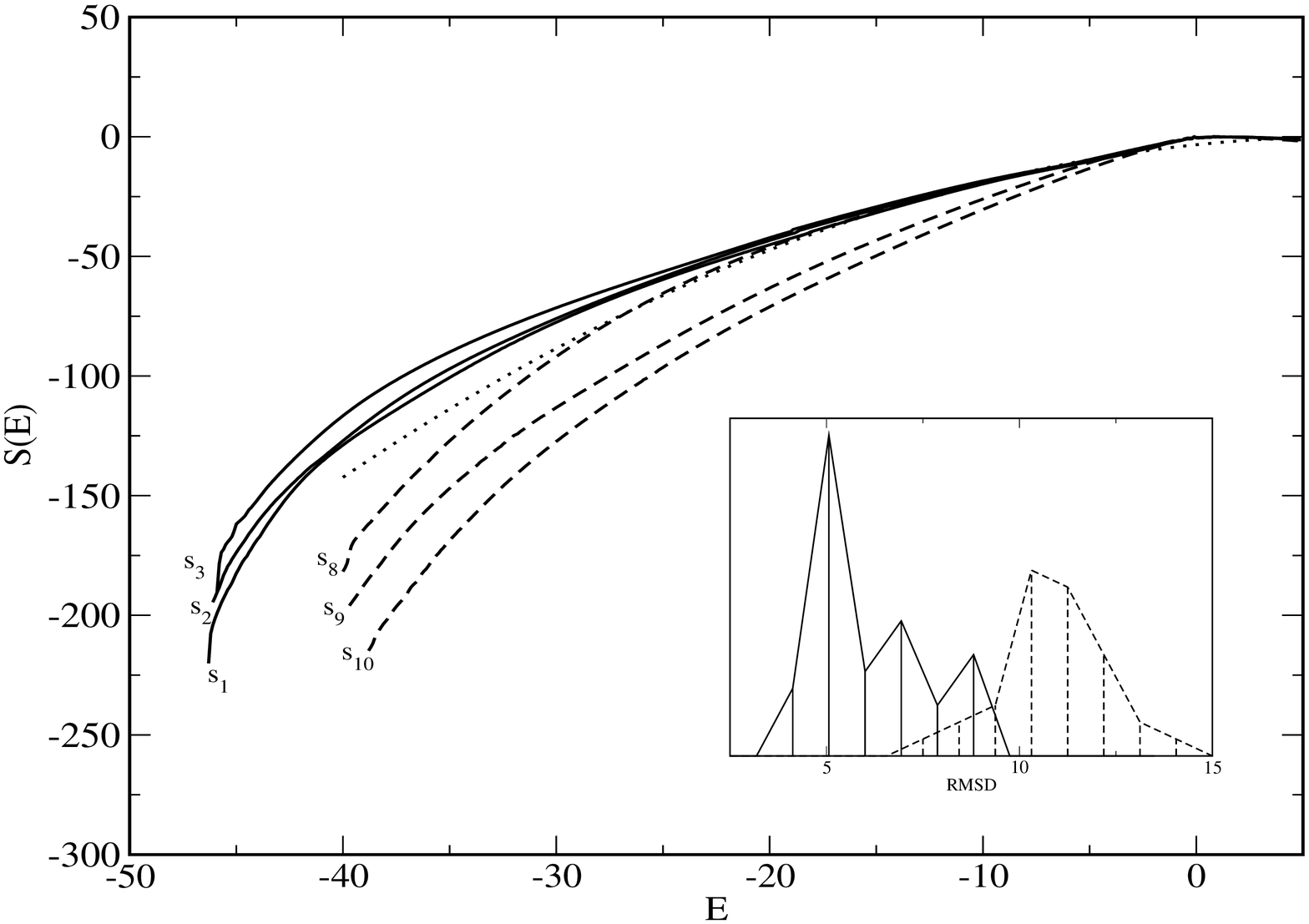,width=12cm}}
\end{figure}

\newpage
\clearpage

\begin{figure}
\centerline{\psfig{file=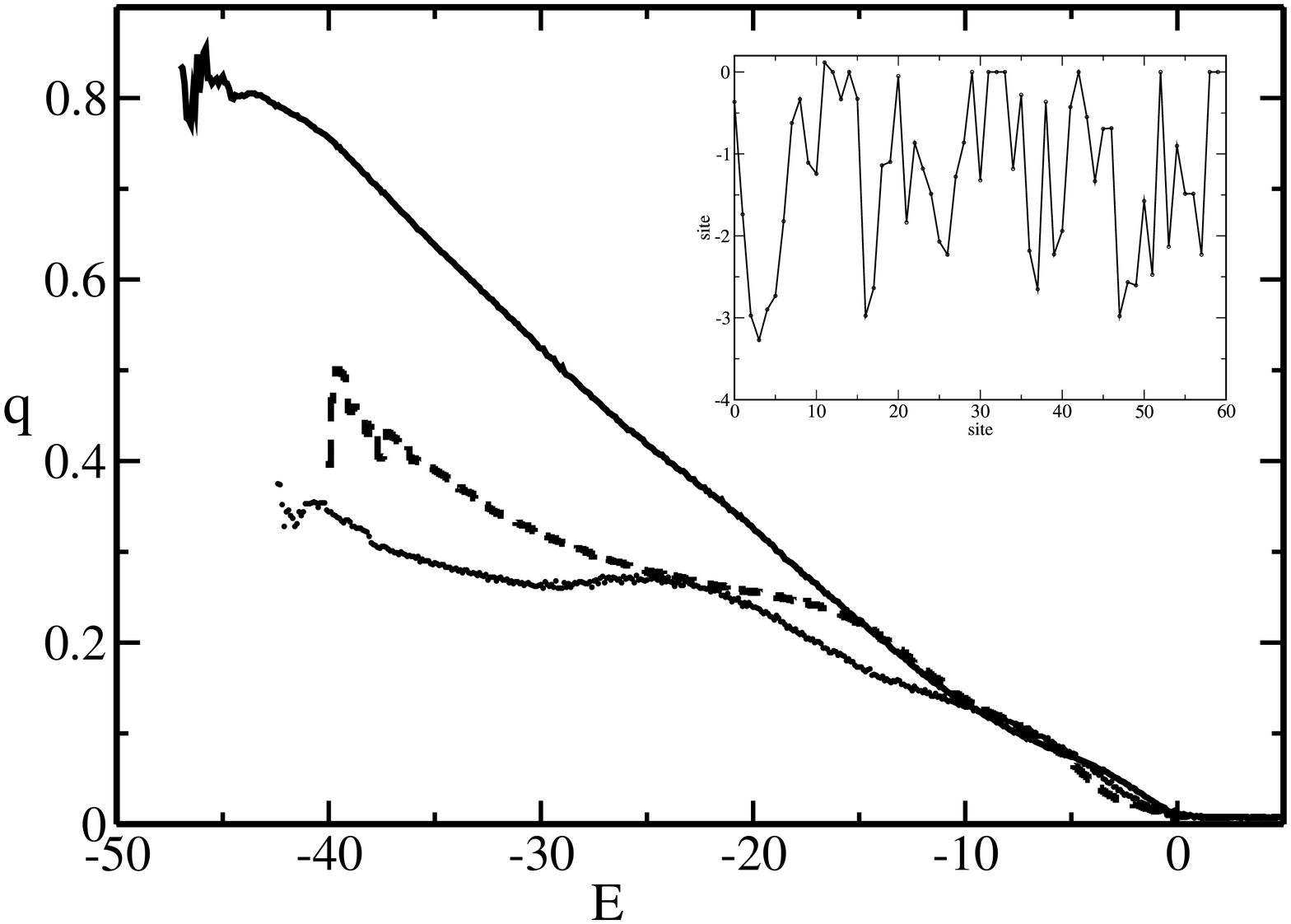,width=12cm}}
\end{figure}

\end{document}